\documentclass[twocolumn,english,english,aps,prb,superscriptaddress,showpacs]{revtex4}
\usepackage{epsfig,graphicx,times}
\newcommand{\mb}[1]{\mathbf{#1}}
\newcommand{\mr}[1]{\mathrm{#1}}
\newcommand{\ket}[1]{\left| #1\right\rangle}   
\newcommand{\bra}[1]{\left\langle #1 \right|} 
\newcommand{\brkt}[3]{\left\langle #1 \Big| #2 \Big| #3\right\rangle}

\newcommand{\Norm}[1]{\left\langle #1 | #1\right\rangle}  
\newcommand{\bigbrace}[1]{\big \{#1 \big\}}
\newcommand{\pt}[1]{\left(#1\right)}  

\newcommand{\Top}{-\frac{1}{2}\nabla^2}
\newcommand{\Hs}{H}

\newcommand{\psim}{\psi^{(m)}}
\newcommand{\psism}{\psi_\sigma^{(m)}}
\newcommand{\zetam}{\zeta_\sigma^{(m)}}
\newcommand{\phim}{\phi^{(m)}}

\newcommand{\vW}{von Weizs\"acker }

\begin{document}

\title{Conjugate-Gradient Optimization Method for Orbital-free Density Functional Calculations}
\author{Hong Jiang}
\affiliation{Department of Chemistry, Duke University, Durham, North Carolina 27708-0354}
\affiliation{Department of Physics, Duke University, Durham, North Carolina 27708-0305}

\author{Weitao Yang}
\email[]{weitao.yang@duke.edu}
\thanks{Corresponding author}
\affiliation{Department of Chemistry, Duke University, Durham, North Carolina 27708-0354}

\date{\today{}}

\begin{abstract}
Orbital-free density functional theory as an extension of traditional Thomas-Fermi theory has attracted a lot of interest in the past decade because of developments in both more accurate kinetic energy functionals and highly efficient numerical numerical methodology. In this paper, we developed a new conjugate-gradient method for the numerical solution of spin-dependent extended Thomas-Fermi equation by incorporating techniques previously used in Kohn-Sham calculations. The key ingredient of the new method is an approximate line-search scheme and a collective treatment of two spin densities in the case of spin-dependent ETF problem. Test calculations for a quartic two-dimensional quantum dot system and a three-dimensional sodium cluster,$\mr{Na}_{216}$, with a local pseudopotential demonstrate that the method is accurate and efficient.  
\end{abstract}
\maketitle

\section{Introduction}
The most attractive feature of density functional theory (DFT) is its use of electron density as the basic variable, \cite{ParrYang89,DreizlerGross90,Jones89RMP} which is established by the Hohenberg-Kohn theorems.\cite{Hohenberg64} The pre-modern orbital-free DFT formalism, mainly the Thomas-Fermi-Dirac-\vW model, \cite{Thomas27, Fermi27, Dirac30, Weizsacker35} though impressive for its simplicity compared to much more complicated wave-function-based approaches, is not quite accurate. Indeed, the great success of modern DFT is due to the introduction of single-particle orbitals in the Kohn-Sham (KS) scheme,\cite{Kohn65} which has become the mainstay of applications of the DFT formalism. The pursuit of more accurate orbital-free density functional theory is nevertheless persistent,\cite{Hohenberg64,Alonso78,Yang86, DePristo87,Lee91,Wang01,Wang92,Perrot94,Brack97,Garcia96,Wang98,Wang99,WangCarter00,Chan01} and has obtained new momentum in recent efforts in {\em ab initio} molecular dynamics simulations of complex systems because of its intrinsic linear scaling behavior. \cite{Shah94,Smargiassi94,Govind94,Blaise97,Foley96,Domps98,Watson00} In this paper, We will use the general term, extended Thomas-Fermi (ETF), to denote all these approaches.

Besides its promising power as a numerical modeling tool, Thomas-Fermi and its extensions also provide the starting point for a lot of other theoretical models.\cite{Wang95,Sinha00,Shi02,Ullmo01b} For instance, in the Strutinsky model \cite{Strutinsky68} of quantum dots, \cite{Ullmo01a,Ullmo01b} Thomas-Fermi theory accounts for classical charging effects, and quantum effects are incorporated by considering the single-particle interference in the Thomas-Fermi effective potential and the residual interaction between oscillating part of electron density. \cite {Ullmo01b}

In all these cases, an efficient numerical method to solve the ETF equation is a necessary ingredient. The ETF equation can be treated either as a nonlinear self-consistent problem, or as a constrained minimization problem. In the former case, a naive implementation of self-consistent iteration is not numerically stable because of the so-called charge-sloshing effect; \cite{Payne92RMP} Elaborate mixing techniques have to be employed to achieve rapid convergence.\cite{Kresse96, Yang86} In the latter case, the simplest approach is the steepest-descent (SD) method, \cite{NumRecip,Payne92RMP, WangCarter00} which, though simple to implement, is a well-known ``poor'' minimization algorithm.\cite{NumRecip} Based on analogy to the Kohn-Sham problem, Wang et al.  formulated the energy minimization in terms of a damped second-order equation.\cite{WangCarter00} The conjugate-gradient (CG) method is one of the most efficient algorithms for numerical optimization, but its implementation to constrained minimization problems is usually very complicated.\cite{WangCarter00} Inspired by a direct minimization conjugated gradient method developed for the Kohn-Sham problem, \cite{Teter89, Payne92RMP,Jiang03b} in this paper, we develop a new CG method to solve the ETF equation, which is simple to implement and very efficient. 

The paper is organized as followings. In the next section, the ETF equation is first cast into an illuminating form, followed by a detailed description of the new CG method. Section III reports some numerical tests in a two-dimensional quantum dot system and  a three-dimensional sodium cluster. The final section summarizes the paper.  

\section{Theory}
\subsection{Formulation of the problem}
According to spin density functional theory,\cite{ParrYang89, DreizlerGross90} the ground state total energy of a $N$-electron interacting system in a local external potential, $V_\mr{ext}(\mb{r})$, is a unique functional of spin densities, $\rho_\sigma(\mb{r})$, under the constraints,
\begin{equation}
\int \rho_\sigma(\mb{r})d\mb{r}=N_\sigma, \label{eq:constraint-rho}
\end{equation}
where $\sigma=\alpha,\beta$ denotes spin-up and down indices, respectively, and $N_\sigma$ is the number of electrons of each spin, with $N_\alpha+N_\beta=N$. The total energy can be written as (atomic units are used through the paper) \cite{Hohenberg64,Kohn65}
\begin{eqnarray}
E_\mr{tot}[\rho_\alpha,\rho_\beta]=T_\mr{s}[\rho_\alpha,\rho_\beta]+\int V_\mr{ext}(\mb{r})\rho(\mb{r})d\mb{r} \nonumber  \\
+\frac{1}{2}\int\int\frac{\rho(\mb{r})\rho(\mb{r'})}{|\mb{r}-\mb{r}'|}d\mb{r}d\mb{r}'+E_\mr{xc}[\rho_\alpha,\rho_\beta], \label{eq:Etot}
\end{eqnarray} 
where $T_s[\rho_\alpha,\rho_\beta]$ is the spin-dependent kinetic energy functional of the fictitious non-interacting system that has the same ground state spin densities as the interacting one, and $E_\mr{xc}[\rho_\alpha,\rho_\beta]$ is the exchange-correlation (XC) energy functionals.     $T_s[\rho_\alpha,\rho_\beta]$ is related to the spin-conpensated kinetic energy functional, $T_s^0[\rho]$, by \cite{Oliver79,ParrYang89}
\begin{equation}
T_s[\rho_\alpha,\rho_\beta]=\frac{1}{2}T_s^0[2\rho_\alpha]+ \frac{1}{2}T_s^0[2\rho_\beta].
\label{eq:Ts-spin}
\end{equation}
In the standard Kohn-Sham scheme, \cite{Kohn65,ParrYang89,DreizlerGross90} $T_s^0$ is calculated by Kohn-Sham single-particle orbitals, which are eigenfunctions of an effective single-particle Hamiltonian. In the orbital-free ETF theory, however, $T_s^0$ is approximated as an explicit functional of the density
\begin{equation}
T_s^0[\rho]=T_\mr{TF}^0[\rho]+\lambda T_\mr{W}^0[\rho]+T_\mr{nl}^0[\rho].\label{eq:E-kin}
\end{equation}
The first term is the Thomas-Fermi kinetic energy functional, a local functional of density, $T_\mr{TF}^0[\rho]=\int t_\mr{TF}(\rho(\mb{r}))d\mb{r}$, where $t_\mr{TF}(\rho)$ is the kinetic energy density of a homogeneous electron system with electron density $\rho$,
\begin{equation}
t_\mr{TF}(\rho)=\left\{ \begin{array}{ll}
\frac{\pi}{2}\rho^2                   \qquad & \textrm{(2D)} \\
\frac{3}{10}(3\pi^2)^{2/3}\rho^{5/3}.  \qquad & \textrm{(3D)}
\end{array} \right.
\end{equation}
$T_\mr{W}^0[\rho]$  is the \vW functional, which is the exact kinetic energy functional in the limit of rapidly varying densities. In both 2D and 3D,  $T_\mr{W}^0[\rho]$ takes the following form \cite{ParrYang89, DreizlerGross90, Wang95}
\begin{equation}
T_\mr{W}^0[\rho]=\frac{1}{8}\int\frac{|\nabla\rho(\mb{r})|^2}{\rho(\mb{r})}d\mb{r}.\label{eq:Tw}
\end{equation}
$\lambda$ is an empirical parameter that is widely used to correct the over-estimation of the \vW term.\cite{ParrYang89} In this paper, we take $\lambda=0.25$. The third term in Eq. (\ref{eq:E-kin}) represents all other non-local extensions that have been the interest of recent efforts.\cite{DePristo87,Lee91,Wang01,Wang92,Brack97,Garcia96,Wang98,Wang99,WangCarter00} Since this paper addresses mainly numerical aspects, we will focus on the so-called TF-$\lambda$W method, which neglects the third term in Eq. (\ref{eq:E-kin}). The techniques developed here are nevertheless general to other more elaborated kinetic energy functional forms. 

Instead of minimizing the total energy directly over the spin densities, we introduce a new quantity, \cite{Govind94,Shah94,Govind95,WangCarter00}
$\psi_\sigma(\mb{r})$, defined by 
\begin{equation}
\rho_\sigma(\mb{r})=(\psi_\sigma(\mb{r}))^2, 
\label{eq:def-psi}
\end{equation}
which can be regarded as quasi-orbitals. The total energy can then be written as the functional of $\psi_\sigma(\mb{r})$, 
\begin{equation}
\widetilde{E}[\psi_\alpha,\psi_\beta] \equiv E_\mr{tot}[(\psi_\alpha(\mb{r}))^2,(\psi_\beta(\mb{r}))^2] 
\end{equation}
and the constraint Eq. (\ref{eq:constraint-rho}) is transformed to the following normalization condition,
\begin{equation}
\int \psi_\sigma(\mb{r})^2 d\mb{r}=N_\sigma,\label{eq:constraint-psi}
\end{equation}
Taking the contraint Eq. (\ref{eq:constraint-psi}) into account by Lagrange's multiplier method, we define 
\begin{equation}
W[\psi_\alpha,\psi_\beta] \equiv  \widetilde{E}[\psi_\alpha,\psi_\beta]- \sum_{\sigma=\alpha,\beta} \widetilde{\mu}_\sigma \Big\{ \int (\psi_\sigma(\mb{r}))^2 d\mb{r}-N_\sigma 
\Big\}
\end{equation} 
where $\widetilde{\mu}_\sigma $ is the chemical potential for spin $\sigma$ electrons. 
 
The advantage of using $\psi_\sigma$ instead of spin densities as the basic variable is two-fold:\cite{Govind94,Shah94,Govind95,WangCarter00} (1) The requirement that $\rho_\sigma(\mb{r})$ has to be positive can be cumbersome to impose in numerical implementations, which is, however, trivial in the case of the $\psi_\sigma$ formulation; (2) more importantly, by introducing $\psi_\sigma$ we can transform the ETF problem to exactly the same form as the KS problem so that all efficient techniques that have been developed in past decades for the KS equations can now be easily extended to the ETF problem. The gradient of the total energy with respect to $\psi_\sigma$ is 
\begin{eqnarray}
\frac{\delta W[\psi_\alpha,\psi_\beta]}{\delta \psi_\sigma(\mb{r})}
&=&\frac{\delta E[\rho_\alpha,\rho_\beta]}{\delta\rho_\sigma(\mb{r})}
   \frac{\partial\rho_\sigma(\mb{r})}{\partial\psi_\sigma(\mb{r})}
   -2\widetilde{\mu}_\sigma \psi_\sigma(\mb{r})\nonumber\\
&=& 2 \Big\{ \frac{\delta E[\rho_\alpha,\rho_\beta]}{\delta\rho_\sigma(\mb{r})}
   -\widetilde{\mu}_\sigma \Big\} \psi_\sigma(\mb{r}) \nonumber \\
&=& 2 \Big\{ \frac{\delta T_\mr{TF}}{\delta 
\rho_\sigma(\mb{r})}+\lambda\frac{\delta T_\mr{W}}{\delta \rho_\sigma(\mb{r})}+V_\mr{eff}^\sigma(\mb{r}) -\widetilde{\mu}_\sigma \Big\} 
\psi_\sigma(\mb{r}), \nonumber 
\\\label{eq:grad-1}
\end{eqnarray}where
\begin{equation}
V_\mr{eff}^\sigma(\mb{r}) \equiv V_\mr{ext}(\mb{r})+\int\frac{\rho(\mb{r}')}{|\mb{r}-\mb{r}'|}d\mb{r}'+\frac{\delta E_\mr{xc}[\rho_\alpha,\rho_\beta]}{\delta\rho_\sigma(\mb{r})}. \label{eq:Veff}
\end{equation}
For $T_\mr{W}$ with the form of Eq. (\ref{eq:Tw}), we have 
\begin{equation}
\frac{\delta T_\mr{W}}{\delta \rho_\sigma(\mb{r})}\psi_\sigma(\mb{r})=\Top\psi_\sigma(\mb{r}),
\end{equation}
from which, Eq. (\ref{eq:grad-1}) can be written concisely as 
\begin{equation}
\frac{\delta W}{\delta \psi_\sigma(\mb{r})}=2\lambda \big\{ \Hs_\sigma 
\psi_\sigma(\mb{r})-\mu_\sigma\psi_\sigma(\mb{r})\}, \label{eq:grad-2}
\end{equation}  
where $\mu_\sigma \equiv \widetilde{\mu}_\sigma/\lambda$, and $\Hs_\sigma $ is defined as 
\begin{equation}
\Hs_\sigma(\mb{r};[\rho_\alpha,\rho_\beta])=\Top+\lambda^{-1}\Big\{\frac{\delta T_\mr{TF}}{\delta \rho_\sigma(\mr{r})}+V_\mr{eff}^\sigma(\mb{r})\Big\}
\label{eq:H}
\end{equation}
The variational principle requires $\delta W/\delta \psi_\sigma(\mb{r})=0$, which 
leads to 
\begin{equation}
\Hs_\sigma \psi_\sigma(\mb{r})=\mu_\sigma\psi_\sigma(\mb{r}), \label{eq:scf-eq}
\end{equation}
which has the same form as the KS equations, but much simplified since there is only one ``orbital'' for each spin state. As in the KS problem, Eq. (\ref{eq:scf-eq}) is a non-linear  equation that has to be solved in a self-consistent way. 

\subsection{A conjugate-gradient method for minimization of ETF total energy}
The formulation of the ETF problem using quasi-orbitals opens up a lot of new possibilities to solve the ETF problem. As in the case of the KS problem, mainly there are two types of approaches, self-consistent and direct minimization methods.\cite{Payne92RMP,Kresse96,Jiang03b} Here we introduced a direct minimization conjugate-gradient (DMCG) method. Such a method for the KS problem is well developed in plane wave \emph{ab initio} modeling of semi-conductor material systems, \cite{Teter89,Payne92RMP}. It was further improved by Jiang et al. in their KS-DFT study of quantum dots. \cite{Jiang03,Jiang03b}

Starting from an initial guess, a conjugate-gradient algorithm for a numerical optimization problem usually involves three steps: \cite{NumRecip,Payne92RMP,Jiang03b}(1) Calculate the steepest descent vector; (2) Construct the conjugate gradient vector; and finally (3) Update the optimization variables  by moving along the conjugate vector direction for a certain distance that is determined  either by an exact line search or by approximations. The complication in the ETF problem is the normalization constraint [Eq. (\ref{eq:constraint-rho}), or (\ref{eq:constraint-psi})]. The algorithm described below is very similar to what we developed for the KS problem, but there are also some subtle differences that are critical for optimal efficiency. We will describe the algorithm for the spin-dependent case, and its application to the spin-independent case is straightforward. 
  
At the $m$-th iteration, the SD vector is calculated as the negative gradient vector [Eq. (\ref{eq:grad-2})] (Dirac's notation for state vectors is used as in Ref [\onlinecite{Jiang03b}]) 
\begin{equation}
\ket{\zetam}=2 \lambda\bigbrace{ \mu_\sigma^{(m)} -\Hs_\sigma^{(m)}} \ket{\psism},\label{eq:zeta}
\end{equation}
with $\mu_\sigma^{(m)}\equiv\bra{\psism} \Hs_\sigma^{(m)} \ket{\psism}/N_\sigma$. The CG vector is then calculated as 
\begin{equation}
\ket{\varphi_\sigma^{(m)}}=\ket{\zeta_\sigma^{(m)}}+\gamma_\sigma^{(m)} \ket{\varphi_\sigma^{(m-1)}}
\end{equation}
with 
\begin{equation}
\gamma_\sigma^{(m)}=\frac{\Norm{\zeta_\sigma^{(m)}}}{\Norm{\zeta_\sigma^{(m-1)}}},
\end{equation}
for $m>1$ and $\gamma_\sigma^{(m)}=0$ for $m=1$. To satisfy the normalization constraint of $\psi_\sigma$, the CG vector is further orthogonalized to $\ket{\psism}$ and normalized to $N_\sigma$,
\begin{eqnarray}
\ket{{\varphi'}_\sigma^{(m)}} =\left( 1-\frac{\ket{\psi_\sigma^{(m)}}
\bra{\psi_\sigma^{(m)}}}{N_\sigma}\right ) \ket{\varphi_\sigma^{(m)}}.\\
\ket{\phi_\sigma^{(m)}}=\ket{{\varphi'}_\sigma^{(m)}} \left (\frac{ N_\sigma}{\Norm{{\varphi'}_\sigma^{(m)}}}\right )^{1/2}
\end{eqnarray}
$\psi_\sigma$ is then updated by 
\begin{equation}
\ket{\psi_\sigma^{(m+1)}}=\ket{\psi_\sigma^{(m)}}\cos\theta_\sigma^\mr{min}+\ket{\phi_\sigma^{(m)}}\sin\theta_\sigma^\mr{min}
\end{equation}
where the values of $\theta_\sigma^\mr{min}$ are determined by minimizing the total energy as a function of $\theta_\alpha$ and $\theta_\beta$,
\begin{eqnarray}
E(\theta_\alpha,\theta_\beta) &\equiv& E_\mr{tot}[\rho_\alpha(\mb{r};\theta_\alpha),\rho_\beta(\mb{r};\theta_\beta)]\nonumber \\
&\equiv &\widetilde{E}[\psi_\alpha(\mb{r};\theta_\alpha),\psi_\beta(\mb{r};\theta_\beta)]
\end{eqnarray}
with 
\begin{equation}
\psi_\sigma(\mb{r};\theta_\sigma)=\psi_\sigma^{(m)}(\mb{r})\cos\theta_\sigma+\phi_\sigma^{(m)}(\mb{r})\sin\theta_\sigma 
\end{equation}
and $\rho_\sigma(\mb{r}; \theta_\sigma)=(\psi_\sigma(\mb{r}; \theta_\sigma))^2$, which is equivalent to minimizing the Lagrangian $W[\psi_\alpha,\psi_\beta]$ since the normalization constraints are intrinsically imposed here. 

The first and second derivatives of $E(\theta_\alpha,\theta_\beta)$ with respect to $\theta_\sigma$ can be obtained by
\begin{eqnarray}
\frac{\partial E(\theta_\alpha,\theta_\beta)}{\partial \theta_\sigma}
&=&\int d\mb{r} \frac{\delta \widetilde{E}}{\delta \psi_\sigma(\mb{r};\theta_\sigma)}
    \frac{\partial\psi_\sigma(\mb{r};\theta_\sigma)}{\partial \theta_\sigma} \nonumber \\
&=& 2\lambda \brkt{\dot{\psi}_\sigma}{H_\sigma(\theta_\alpha,\theta_\beta)}{\psi_\sigma(\theta_\sigma)}
\label{eq:1-deriv}
\end{eqnarray}
and 
\begin{eqnarray}
\frac{\partial^2E(\theta_\alpha,\theta_\beta)}
{\partial \theta_\sigma\partial\theta_{\sigma'}}= 2\lambda&\Bigg\{&
\brkt{\ddot{\psi}_\sigma(\theta_\sigma)}{H_\sigma(\theta_\alpha,\theta_\beta)}{\psi_\sigma(\theta_\sigma)}\delta_{\sigma,\sigma'}\nonumber \\
&+&\brkt{\dot{\psi}_\sigma(\theta_\sigma)}{H_\sigma(\theta_\alpha,\theta_\beta)}{\dot{\psi}_\sigma(\theta_\sigma)} \delta_{\sigma,\sigma'} \nonumber\\
&+&\brkt{\dot{\psi}_\sigma(\theta_\sigma)}{\frac{\partial H_\sigma(\theta_\alpha,\theta_\beta)}{\partial\theta_{\sigma'}}}{\psi_\sigma(\theta_\sigma)}\Bigg\}\nonumber\\
\end{eqnarray}
with 
\begin{equation}
H_\sigma(\theta_\alpha,\theta_\beta) \equiv H(\mb{r};[\rho_\alpha(\theta_\alpha),\rho_\beta(\theta_\beta)]) 
\end{equation}
\begin{eqnarray}
\frac{\partial H_\sigma(\theta_\alpha,\theta_\beta)}{\partial\theta_{\sigma'}}
=\lambda^{-1}\Big\{ \big( 2t_\mr{TF}''(2\rho_\sigma(\theta_\sigma))\delta_{\sigma,\sigma'} 
\nonumber \\
+\frac{\delta^2E_\mr{XC}[\rho_\alpha,\rho_\beta]}{\delta \rho_\sigma \delta \rho_{\sigma'}} \big)\dot{\rho}_{\sigma'}(\theta_\sigma') + \int d\mb{r}'\frac{\dot{\rho}_{\sigma'}(\mb{r}';\theta_\sigma')}{|\mb{r}-\mb{r}'|}\Big\} 
\end{eqnarray}
\begin{eqnarray}
\dot{\psi}_\sigma(\theta_\sigma) 
   &\equiv& \frac{\partial \psi_\sigma(\theta_\sigma)}{\partial\theta_\sigma}
   = -\psi_\sigma^{(m)}\sin\theta_\sigma+\phi_\sigma^{(m)}\cos\theta_\sigma \\
\ddot{\psi}_\sigma(\theta_\sigma)
   &\equiv& \frac{\partial^2 \psi_\sigma(\theta_\sigma)}{\partial\theta_\sigma^2}
   =-\psi_\sigma(\theta_\sigma) \\
\dot{\rho}_\sigma(\theta_\sigma)
   &\equiv&\frac{\partial \rho_\sigma(\theta_\sigma)}{\partial\theta_\sigma}
   =2\psi_\sigma(\theta_\sigma)\dot{\psi}_\sigma(\theta_\sigma).
\end{eqnarray}

There are a lot of standard optimization techniques available to solve this two-variable minimization problem (single-variable minimization for the spin-independent ETF). \cite{NumRecip,Bertsekas99} We will, however, pursue an approximate scheme similar to the method we used in the KS case.\cite{Jiang03b} We first transform Eq. (\ref{eq:1-deriv}) into a more illuminating form
\begin{eqnarray}
\frac{\partial E(\theta_\alpha,\theta_\beta)}{\partial \theta_\sigma }
=2\lambda\bra{-\psi_\sigma^{(m)}\sin\theta_\sigma+\phi_\sigma^{(m)}\cos\theta_\sigma } \nonumber \\
\times H_\sigma(\theta_\alpha,\theta_\beta) \ket{\psi_\sigma^{(m)}\cos\theta_\sigma+\phi_\sigma^{(m)}\sin\theta_\sigma } \nonumber\\
=\lambda \big ( -A_\sigma(\theta_\alpha,\theta_\beta) \sin 2\theta_\sigma+ B_\sigma(\theta_\alpha,\theta_\beta) 
\cos2\theta_\sigma) \label{eq:d1-2}
\end{eqnarray}
with
\begin{eqnarray}
A_\sigma(\theta_\alpha,\theta_\beta)\equiv & &\bra{\psi_\sigma^{(m)}} \Hs_\sigma(\theta_\alpha,\theta_\beta)\ket{\psi_\sigma^{(m)}} \nonumber\\
&-&\bra{\phi_\sigma^{(m)}}\Hs_\sigma(\theta_\alpha,\theta_\beta)\ket{\phi_\sigma^{(m)}}
\end{eqnarray}
and
\begin{equation}
B_\sigma(\theta_\alpha,\theta_\beta)\equiv 2\bra{\phi_\sigma^{(m)}} \Hs_\sigma(\theta_\alpha,\theta_\beta)\ket{\psi _\sigma^{(m)}},
\end{equation} 
where we have used the fact that both $\psi_\sigma$ and $\phi_\sigma$ are real. Now we introduce the approximation
\begin{equation}
\Hs_\sigma(\theta_\alpha,\theta_\beta) \simeq \Hs_\sigma(0,0), \label{eq:approx}
\end{equation}
so that by setting the first derivate, Eq. (\ref{eq:d1-2}), to be zero, we obtain 
\begin{equation} 
\theta_\sigma^\mathrm{min} \simeq \frac{1}{2}\tan ^{-1}\frac{B_\sigma(0,0)}{A_\sigma(0,0)}. 
\label{eq:linmin0}
\end{equation}
The validity of Eq. (\ref{eq:approx}) can be established by the following observations: The dominant parts of the effective Hamiltonian $H_\sigma$ are the kinetic energy operator and external potential, which are independent of $\theta_\sigma$[See Eq.(\ref{eq:H}]; the other parts of $H_\sigma$ depend on $\theta_\sigma$ through $\rho_\sigma(\theta_\sigma) $, which plays a weaker role in determining $\theta_\sigma^\mr{min}$. In the Appendix, we formulate a more accurate approximation that goes one step beyond Eq.(\ref{eq:approx}), in which the dependence of the Hatree potential on $\theta_\sigma$ is incorporated.  

The algorithm described above is very close to the band-by-band DMCG method used in Kohn-Sham calculations.\cite{Teter89,Payne92RMP,Jiang03b} There is, however, one critical difference. In the spirit of the band-by-band scheme, one might treat two spin densities alternatively: Iterate only one spin density at a time for $N_\mr{band}$ times with the other spin density fixed. Such a sequential treatment has the disadvantage that the conjugate gradient relaxation is disrupted every $N_\mr{band}$ times, which impairs partly the high-efficiency intrinsic to a conjugate-gradient algorithm. We also found in our numerical tests that the approximate line-minimization scheme using Eq. (\ref{eq:linmin0}) is sometimes numerically unstable, and a more accurate approximation to $\theta^\mr{min}$ like that described in Appendix or an ``exact'' line search is required. In spite of that, the sequential scheme is still much more efficient than the steepest descent method; When an exact line search is required, Eq. (\ref{eq:linmin0}) provides a very accurate initial guess. We will denote that treatment as the SCG (for sequential conjugate-gradient) method in the following section. In contrast, the method described above iterates the two spin densities simultaneously, and in doing so has taken full advantage of the fact that the two quasi spin orbitals are not required to be orthogonal to each other. In this treatment, the approximation made in Eq. (\ref{eq:linmin0}) is found to be stable and accurate, which dramatically reduces the computation efforts; iterations are always carried out in the conjugate gradient direction. Because the normalization constraint is imposed in each iteration, there is no accumulation of numerical errors that may occur in some CG algorithms. We will denote this treatment as CCG (for concurrent conjugate-gradient). For the spin-independent case, these two approaches are identical. 

\section{Numerical Tests}
We will report numerical results only for finite systems, but conclusions from these test calculations should be applicable to periodic infinite systems as well. For other important components of a full implementation of the ETF theory to real systems, we essentially use same techniques as we did in the KS case: \cite{Jiang03b} We use the particle-in-the-box basis for the representation of quasi-orbitals, $\psi_\sigma$, which is a variant of plane-wave basis for finite systems; The action of the effective Hamiltonian on quasi-orbitals, $H_\sigma \psi_\sigma$ is effected by fast-sine transform; \cite{Jiang03b} For the Hartree potential, we use the Fourier convolution method on a doubly extended grid. \cite{Martyna99,Jiang03b}  Local spin density approximation is used for $E_\mr{xc}$. In particular, we use the Tanatar-Ceperley functional \cite{Tanatar89} for 2D systems and  the Vosko-Wilk-Nusair functional \cite{Vosko80} for 3D systems.  

We first test the performance of the new method in a 2D quantum dot model system with a coupled quartic oscillator potential (QOP), \cite{Bohigas93,Jiang03}
\begin{equation}
\label{eq:qop}
V_\mathrm{ext}(x,y)=a\left( \frac{x^{4}}{b}+by^{4}-2\lambda x^{2}y^{2}+
\gamma (x^{2}y-xy^{2})r\right),
\end{equation}
with \( r=\sqrt{x^2+y^2}\), $a=10^{-4}$, $b=\pi/4$, $\lambda=0.53$ and $\gamma=0.2$. This potential was used by the authors in numerical studies of e-e interaction effects in quantum dots.\cite{Jiang03,Jiang03b} The numerical results are obtained for $N=200$. For the spin-dependent case, we consider the triplet state, i.e. $N_\alpha=101$ and  $N_\beta=99$. 

\begin{figure}
\includegraphics[width=2.8in,clip]{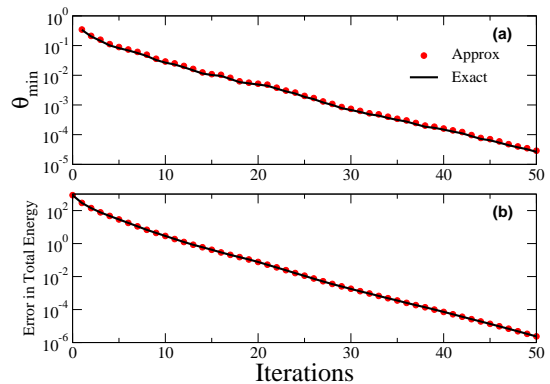}
\caption{\label{fig:linmin} 
Comparison of approximate and exact line search methods in spin-independent ETF. (a) $\theta_\mr{min}$ calculated by the approximate method, Eq. (\ref{eq:linmin0}) (dot), and by exact Brent's method in a typical ETF calculation (line); (b) Convergence errors vs iteration number using approximate and exact line search methods. 
}
\end{figure}

\begin{figure}
\includegraphics[width=2.8in,clip]{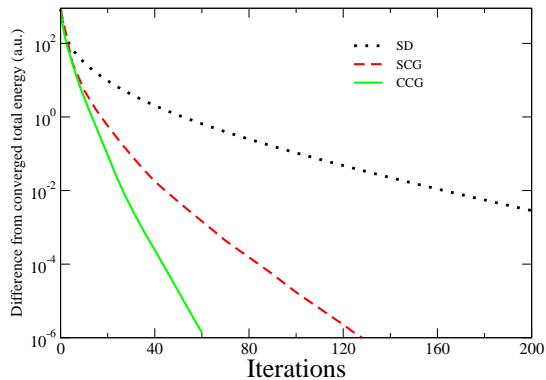}
\caption{\label{fig:conv-cmp}(Color online) Error in the total energy vs iteration number in three different minimization methods for spin-dependent ETF: steepest descent (dot), sequential conjugate-gradient (dash) and collective conjugate gradient (solid). In SCG, Brent's method for line minimization is used, and $N_\mr{band}=5$. }
\end{figure}
\begin{figure}
\includegraphics[width=2.5in,clip]{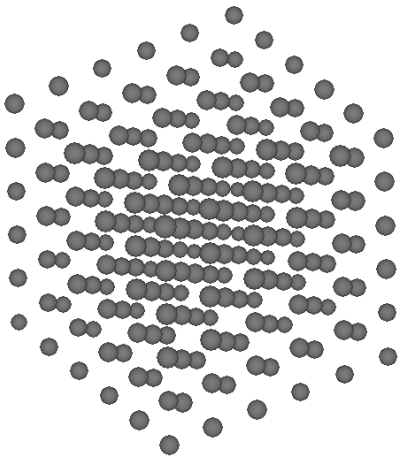}
\caption{\label{fig:geom}The geometry of the Na cluster used in the test calculations}
\end{figure}

\begin{figure}
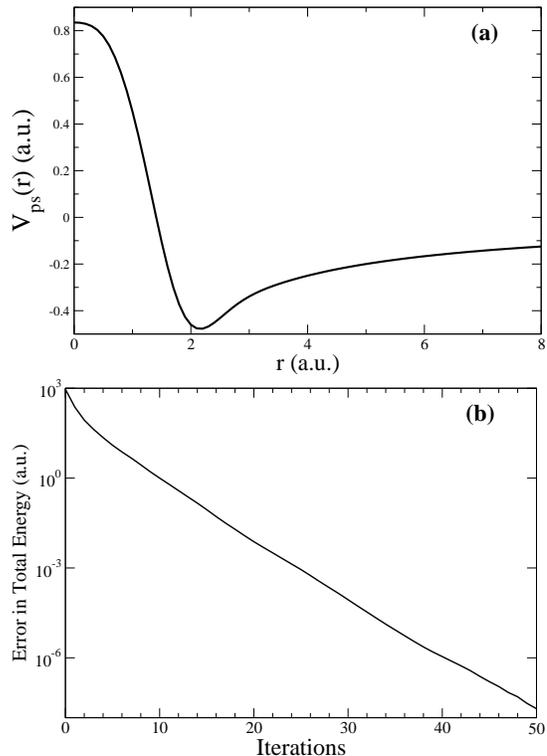

\includegraphics[width=2.8in,clip]{Vps-Na.eps}
\includegraphics[width=2.8in,clip]{Na216-conv-err.eps}
\caption{\label{fig:Na}(a) Local ion pseudopotential for Na used in test calculations of $\mr{Na}_{216}$; (b) Error in the total energy in the $\mr{Na}_{216}$ system vs iteration number.}
\end{figure}
We first test the accuracy of the approximation of Eq. (\ref{eq:linmin0}) by comparing with the exact line search in the spin-independent ETF case. The exact $\theta^\mr{min}$ is calculated by the standard  Brent's method,\cite{NumRecip} in which Eq. (\ref{eq:linmin0}) is used to calculate the initial guess. Fig.\ref{fig:linmin}(a) shows approximate and exact $\theta^\mr{min}$ in a typical minimization process. The approximate $\theta^\mr{min}$ remains in good agreement with the exact one through the iterations. Fig. \ref{fig:linmin}(b) plots the convergence errors vs iteration number using approximate and exact line minimization methods, respectively. The agreement between them is almost perfect. 

Fig. \ref{fig:conv-cmp} shows the convergence behaviors of three different minimization methods, SD, SCG and CCG, in the case of the spin-dependent ETF calculations. For the SD method, we use the formalism in Ref. [\onlinecite{WangCarter00}], but with a exact line search. Using the approximation in Eq. (\ref{eq:linmin0}), the computation cost for a single step in CCG is much less than that of SD and SCG, yet the convergence is still faster to attain in CCG than in SCG as well as in SD. The combination of these two improving factors reduces the computation efforts in CCG by almost two order of magnitude with respect to SCG. 

Finally, to test the performance of our new method in more realistic systems, we consider a  sodium cluster , $\mr{Na}_{216}$. Since our main purpose here is to test the performance of our new method, we take the geometry of the cluster as a cubic without relaxation, as illustrated in Fig. \ref{fig:geom}; the distance between neighboring Na atoms is taken as $4$ atomic units. In this case, the external potential is formed by the superposition of the local pseudo-potential for the Na ion on different sites,
\begin{equation}
V_\mr{ext}(\mb{r})=\sum_i V_\mr{ps}(|\mb{r}-\mb{R}_i|). \label{eq:Vext-Na9}
\end{equation}
For $ V_\mr{ps}(r) $ we use Bachelet-Hamann-Schl\"uter's norm-conserving pseudopotential \cite{Bachelet82} neglecting the non-local part, which is illustrated in Fig. \ref{fig:Na}(a). In the test calculations, we use a box of $28.0^3$ in real space with $81$ mesh points in each direction. We take both the change in the total energy between successive iterations, $\Delta E \equiv E^{(m)}-E^{(m-1)}$, and the norm of the gradient as the convergence criteria. 

Fig. \ref{fig:Na}(b) shows the convergence behavior of our method in this realistic system. As in the previous 2D model system, the convergence can be easily attained in about 50 iterations, which confirms the high efficiency of our new conjugate-gradient method in realistic molecular systems. 

\section{Summary and Discussion}
In this paper, an efficient conjugate gradient  method for direct minimization of the extended Thomas-Fermi total energy  is proposed. The new method  is inspired  by a similar approach for the Kohn-Sham problem. The key ingredient of the new method is an approximate line-search scheme and a collective treatment of two spin densities in the case of spin-dependent ETF problem. The high performance of the new method was verified in a simple 2D model system and a realistic sodium cluster. 

We close the paper with two comments. First, though the method presented in the paper is based on transforming the ETF problem into a KS-like form, which is enabled by  the presence of the \vW term, our method has more general significance in terms of its mathematical structure. We note that the validity of the formulation in Section II.B does not depends on the exact form of the effective Hamiltonian, $H_\sigma$. By defining  
\begin{equation}
\lambda H_\sigma\psi_\sigma \equiv \frac{1}{2} \frac{\delta\widetilde{E}}{\delta\psi_\sigma},
\end{equation}
our method can be easily generalized to other orbital-free DFT formalisms with or without the \vW term.

Second, recently in a benchmark ETF studies on atomic and diatomic systems using Gaussian basis sets, Chan et al. found that all  gradient-based methods including CG and quasi-Newton methods perform poorly in minimizing ETF total energy.\cite{Chan01} Though we tested our new method only in the plane-wave type representation, the formulation of the method is nevertheless general, and therefore should be equally applicable to local basis represented systems. We will leave further investigations to future studies. 

\begin{acknowledgements}
We thank Harold U. Baranger for helpful discussions and a careful reading of the manuscript. This work was supported in part by NSF Grant No. DMR-0103003. 
\end{acknowledgements}

\appendix*
\section{}
The approximation in Eq. (\ref{eq:linmin0}) can be improved by considering the dependence of the Hartree potential on $\theta$. To simplify the notation, we present the formulation only for the spin-independent case, and the spin index is therefore dropped. Instead of neglecting the dependence of $H$ on $\theta$ completely [Eq.(\ref{eq:approx})], we make a less dramatic approximation,
\begin{equation} \label{eq:Happrox2}
H(\mb{r};\theta) \simeq H(\mb{r};0)+\int \frac{\Delta\rho(\mb{r}';\theta)}{|\mb{r}-\mb{r}'|}d\mb{r}'   
\end{equation}
where $\Delta\rho(\mb{r};\theta)$ is the change of the density
\begin{eqnarray} \label{eq:drho}
\Delta\rho(\mb{r};\theta) 
   &\equiv& \rho(\mb{r};\theta)-\rho(\mb{r};0) \nonumber\\
   &=& \left( \psim(\mb{r}) \cos\theta+(\mb{r})\sin\theta \right)^2 -\left(\psim(\mb{r}))\right)^2. \nonumber \\
   &=& -\chi_1(\mb{r}) \sin^2\theta +\chi_2(\mb{r})\cos\theta\sin\theta
\end{eqnarray}
with
\begin{eqnarray}
 \chi_1(\mb{r})&\equiv&\pt{\psim(\mb{r})}^2-\pt{\phim(\mb{r})}^2, \\
 \chi_2(\mb{r})&\equiv&2\psim(\mb{r})\phim(\mb{r}).
\end{eqnarray}
Using Eqs.(\ref{eq:Happrox2},\ref{eq:drho}), we can obtain, after some straightforward transformation, 
\begin{eqnarray}
A(\theta)&=&A_0-C_{11}\sin^2\theta+C_{12}\sin\theta\cos\theta \\
B(\theta)&=&B_0-C_{12}\sin^2\theta+C_{22}\sin\theta\cos\theta
\end{eqnarray} 
with 
\begin{eqnarray}
A_0 &\equiv& \bra{\psim}H(0)\ket{\psim}-  \bra{\phim}H(0)\ket{\phim}, \\
B_0 &\equiv& 2\bra{\phim}H(0)\ket{\psim},  \\
C_{ij}&\equiv&\int \int \frac{\chi_i(\mb{r})\chi_j(\mb{r}')}{|\mb{r}-\mb{r}'|}d\mb{r}d\mb{r}'
,\quad \textrm{for~} i,j= 1, 2. 
\end{eqnarray}
$\theta^\mr{min}$ is then determined by requiring 
\begin{eqnarray} 
\frac{\delta E(\theta)}{\delta \theta} &\simeq& \lambda \bigbrace{-(A_0-\frac{C_{11}}{2})\sin 2\theta+(\frac{B_0-C_{12}}{2})\cos 2\theta \nonumber \\
&-&\frac{C_{11}-C_{22}}{4}\sin 4\theta+\frac{C_{12}}{2}\cos 4\theta},  \label{eq:dE}
\end{eqnarray}
to be zero. Though there is no simple close expression for the root of Eq.(\ref{eq:dE}) like Eq. (\ref{eq:linmin0}), it can be easily solved numerically using standard techniques like the Newton-Raphson method. \cite{NumRecip} The increase in the computation efforts using this more accurate line-minimization approximation is marginal when compared to that using Eq.(\ref{eq:linmin0}). 

\begin{figure}
\includegraphics[width=2.8in,clip]{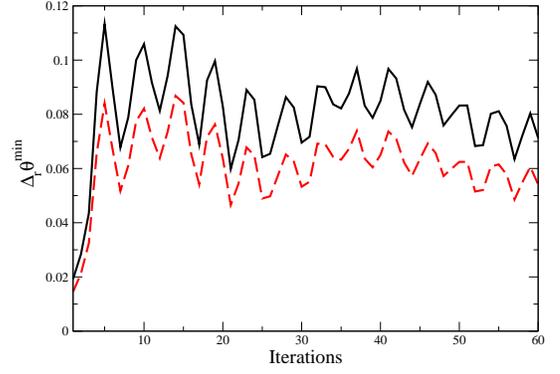}
\caption{\label{fig:linmin-cmp} 
Relative errors in $\theta^\mr{min}$ [Eq. (\ref{eq:theta-err}) using Eq. (\ref{eq:linmin0}) (solid) and Eq. (\ref{eq:dE}) (dashed), respectively, in the spin-independent ETF calculation of the QOP system with $N=200$. The exact  $\theta^\mr{min}$ is calculated by the standard Brent's method.\cite{NumRecip}
}
\end{figure}

We tested the new line minimization method in the QOP system with $N=200$ in the spin-independent case. Fig. (\ref{fig:linmin-cmp}) shows the relative errors in $\theta^\mr{min}$
\begin{equation}\label{eq:theta-err}
\Delta_r\theta^\mr{min} \equiv \frac{\theta^\mr{min}_\mr{approx}-\theta^\mr{min}_\mr{exact}}{\theta^\mr{min}_\mr{exact}}
\end{equation}
during a typical calculation, which shows that the new line-minimization method based on Eq. (\ref{eq:dE}) significantly improved the accuracy of the approximated $\theta^\mr{min}$. On the other hand, the number of iterations required to achieve convergence remains same, which confirms further the validity of Eq.(\ref{eq:linmin0}). We note, however, that the improvement of Eq. (\ref{eq:dE}) with respect to Eq.(\ref{eq:linmin0}) is important to in the SCG scheme for spin-dependent ETF calculations to guarantee the numerical stability, as discussed in the paper. We expect that the improved method will also be useful in systems where Eq.(\ref{eq:linmin0}) might fail.

\bibliography{nano,rmt,dft,method}

\begin{thebibliography}{49}
\expandafter\ifx\csname natexlab\endcsname\relax\def\natexlab#1{#1}\fi
\expandafter\ifx\csname bibnamefont\endcsname\relax
  \def\bibnamefont#1{#1}\fi
\expandafter\ifx\csname bibfnamefont\endcsname\relax
  \def\bibfnamefont#1{#1}\fi
\expandafter\ifx\csname citenamefont\endcsname\relax
  \def\citenamefont#1{#1}\fi
\expandafter\ifx\csname url\endcsname\relax
  \def\url#1{\texttt{#1}}\fi
\expandafter\ifx\csname urlprefix\endcsname\relax\def\urlprefix{URL }\fi
\providecommand{\bibinfo}[2]{#2}
\providecommand{\eprint}[2][]{\url{#2}}

\bibitem[{\citenamefont{Parr and Yang}(1989)}]{ParrYang89}
\bibinfo{author}{\bibfnamefont{R.~G.} \bibnamefont{Parr}} \bibnamefont{and}
  \bibinfo{author}{\bibfnamefont{W.}~\bibnamefont{Yang}},
  \emph{\bibinfo{title}{Density-Functional Theory of Atoms and Molecules}}
  (\bibinfo{publisher}{Oxford University Press}, \bibinfo{address}{New York},
  \bibinfo{year}{1989}).

\bibitem[{\citenamefont{Dreizler and Gross}(1990)}]{DreizlerGross90}
\bibinfo{author}{\bibfnamefont{R.~M.} \bibnamefont{Dreizler}} \bibnamefont{and}
  \bibinfo{author}{\bibfnamefont{E.~K.~U.} \bibnamefont{Gross}},
  \emph{\bibinfo{title}{Density Functional Theory : An Approach to the Quantum
  Many-Body Problem}} (\bibinfo{publisher}{Springer-Verlag},
  \bibinfo{address}{Berlin}, \bibinfo{year}{1990}).

\bibitem[{\citenamefont{Jones and Gunnarsson}(1989)}]{Jones89RMP}
\bibinfo{author}{\bibfnamefont{R.~O.} \bibnamefont{Jones}} \bibnamefont{and}
  \bibinfo{author}{\bibfnamefont{O.}~\bibnamefont{Gunnarsson}},
  \bibinfo{journal}{Rev. Mod. Phys.} \textbf{\bibinfo{volume}{61}},
  \bibinfo{pages}{689} (\bibinfo{year}{1989}).

\bibitem[{\citenamefont{Hohenberg and Kohn}(1964)}]{Hohenberg64}
\bibinfo{author}{\bibfnamefont{P.}~\bibnamefont{Hohenberg}} \bibnamefont{and}
  \bibinfo{author}{\bibfnamefont{W.}~\bibnamefont{Kohn}},
  \bibinfo{journal}{Phys. Rev. B} \textbf{\bibinfo{volume}{136}},
  \bibinfo{pages}{864} (\bibinfo{year}{1964}).

\bibitem[{\citenamefont{Thomas}(1927)}]{Thomas27}
\bibinfo{author}{\bibfnamefont{L.~H.} \bibnamefont{Thomas}},
  \bibinfo{journal}{Proc. Camb. Phil. Soc.} \textbf{\bibinfo{volume}{23}},
  \bibinfo{pages}{542} (\bibinfo{year}{1927}).

\bibitem[{\citenamefont{Fermi}(1927)}]{Fermi27}
\bibinfo{author}{\bibfnamefont{E.}~\bibnamefont{Fermi}},
  \bibinfo{journal}{Rend. Accad. Nazl. Lincei} \textbf{\bibinfo{volume}{6}},
  \bibinfo{pages}{602} (\bibinfo{year}{1927}).

\bibitem[{\citenamefont{Dirac}(1930)}]{Dirac30}
\bibinfo{author}{\bibfnamefont{P.~A.~M.} \bibnamefont{Dirac}},
  \bibinfo{journal}{Proc. Camb. Phil. Soc.} \textbf{\bibinfo{volume}{26}},
  \bibinfo{pages}{376} (\bibinfo{year}{1930}).

\bibitem[{\citenamefont{Weizs\"acker}(1935)}]{Weizsacker35}
\bibinfo{author}{\bibfnamefont{L.~H.} \bibnamefont{Weizs\"acker}},
  \bibinfo{journal}{Proc. Camb. Phil. Soc.} \textbf{\bibinfo{volume}{23}},
  \bibinfo{pages}{542} (\bibinfo{year}{1935}).

\bibitem[{\citenamefont{Kohn and Sham}(1965)}]{Kohn65}
\bibinfo{author}{\bibfnamefont{W.}~\bibnamefont{Kohn}} \bibnamefont{and}
  \bibinfo{author}{\bibfnamefont{L.~J.} \bibnamefont{Sham}},
  \bibinfo{journal}{Phys. Rev. A} \textbf{\bibinfo{volume}{140}},
  \bibinfo{pages}{1133} (\bibinfo{year}{1965}).

\bibitem[{\citenamefont{Alonso and Girifalco}(1978)}]{Alonso78}
\bibinfo{author}{\bibfnamefont{J.~A.} \bibnamefont{Alonso}} \bibnamefont{and}
  \bibinfo{author}{\bibfnamefont{L.~A.} \bibnamefont{Girifalco}},
  \bibinfo{journal}{Phys. Rev. B} \textbf{\bibinfo{volume}{17}},
  \bibinfo{pages}{3735} (\bibinfo{year}{1978}).

\bibitem[{\citenamefont{Yang}(1986)}]{Yang86}
\bibinfo{author}{\bibfnamefont{W.}~\bibnamefont{Yang}}, \bibinfo{journal}{Phys.
  Rev. A} \textbf{\bibinfo{volume}{34}}, \bibinfo{pages}{4575}
  (\bibinfo{year}{1986}).

\bibitem[{\citenamefont{DePristo and Kress}(1987)}]{DePristo87}
\bibinfo{author}{\bibfnamefont{A.~E.} \bibnamefont{DePristo}} \bibnamefont{and}
  \bibinfo{author}{\bibfnamefont{J.~D.} \bibnamefont{Kress}},
  \bibinfo{journal}{Phys. Rev. A} \textbf{\bibinfo{volume}{35}},
  \bibinfo{pages}{438} (\bibinfo{year}{1987}).

\bibitem[{\citenamefont{Lee et~al.}(1991)\citenamefont{Lee, Lee, and
  G.}}]{Lee91}
\bibinfo{author}{\bibfnamefont{H.}~\bibnamefont{Lee}},
  \bibinfo{author}{\bibfnamefont{C.}~\bibnamefont{Lee}}, \bibnamefont{and}
  \bibinfo{author}{\bibfnamefont{P.~R.} \bibnamefont{G.}},
  \bibinfo{journal}{Phys. Rev. A} \textbf{\bibinfo{volume}{44}},
  \bibinfo{pages}{768} (\bibinfo{year}{1991}).

\bibitem[{\citenamefont{Wang et~al.}(2001)\citenamefont{Wang, Stott, and von
  Barth}}]{Wang01}
\bibinfo{author}{\bibfnamefont{B.}~\bibnamefont{Wang}},
  \bibinfo{author}{\bibfnamefont{M.~J.} \bibnamefont{Stott}}, \bibnamefont{and}
  \bibinfo{author}{\bibfnamefont{U.}~\bibnamefont{von Barth}},
  \bibinfo{journal}{Phys. Rev. A} \textbf{\bibinfo{volume}{63}},
  \bibinfo{pages}{052501} (\bibinfo{year}{2001}).

\bibitem[{\citenamefont{Wang and Teter}(1992)}]{Wang92}
\bibinfo{author}{\bibfnamefont{L.~W.} \bibnamefont{Wang}} \bibnamefont{and}
  \bibinfo{author}{\bibfnamefont{M.~P.} \bibnamefont{Teter}},
  \bibinfo{journal}{Phys. Rev. B} \textbf{\bibinfo{volume}{45}},
  \bibinfo{pages}{13196} (\bibinfo{year}{1992}).

\bibitem[{\citenamefont{Perrot}(1994)}]{Perrot94}
\bibinfo{author}{\bibfnamefont{F.}~\bibnamefont{Perrot}}, \bibinfo{journal}{J.
  Phys.:Condens. Matter} \textbf{\bibinfo{volume}{6}}, \bibinfo{pages}{431}
  (\bibinfo{year}{1994}).

\bibitem[{\citenamefont{Brack and Bhaduri}(1997)}]{Brack97}
\bibinfo{author}{\bibfnamefont{M.}~\bibnamefont{Brack}} \bibnamefont{and}
  \bibinfo{author}{\bibfnamefont{R.~K.} \bibnamefont{Bhaduri}},
  \emph{\bibinfo{title}{Semiclassical Physics}}
  (\bibinfo{publisher}{Addison-Wesley}, \bibinfo{address}{Reading, USA},
  \bibinfo{year}{1997}).

\bibitem[{\citenamefont{Garcia-Gonzalez
  et~al.}(1996)\citenamefont{Garcia-Gonzalez, Alvarellos, and
  Chacon}}]{Garcia96}
\bibinfo{author}{\bibfnamefont{P.}~\bibnamefont{Garcia-Gonzalez}},
  \bibinfo{author}{\bibfnamefont{J.~E.} \bibnamefont{Alvarellos}},
  \bibnamefont{and} \bibinfo{author}{\bibfnamefont{E.}~\bibnamefont{Chacon}},
  \bibinfo{journal}{Phys. Rev. A} \textbf{\bibinfo{volume}{54}},
  \bibinfo{pages}{1897} (\bibinfo{year}{1996}).

\bibitem[{\citenamefont{Wang et~al.}(1998)\citenamefont{Wang, Govind, and
  Carter}}]{Wang98}
\bibinfo{author}{\bibfnamefont{Y.~A.} \bibnamefont{Wang}},
  \bibinfo{author}{\bibfnamefont{N.}~\bibnamefont{Govind}}, \bibnamefont{and}
  \bibinfo{author}{\bibfnamefont{E.~A.} \bibnamefont{Carter}},
  \bibinfo{journal}{Phys. Rev. B} \textbf{\bibinfo{volume}{58}},
  \bibinfo{pages}{13465} (\bibinfo{year}{1998}).

\bibitem[{\citenamefont{Wang et~al.}(1999)\citenamefont{Wang, Govind, and
  Carter}}]{Wang99}
\bibinfo{author}{\bibfnamefont{Y.~A.} \bibnamefont{Wang}},
  \bibinfo{author}{\bibfnamefont{N.}~\bibnamefont{Govind}}, \bibnamefont{and}
  \bibinfo{author}{\bibfnamefont{E.~A.} \bibnamefont{Carter}},
  \bibinfo{journal}{Phys. Rev. B} \textbf{\bibinfo{volume}{60}},
  \bibinfo{pages}{16350} (\bibinfo{year}{1999}).

\bibitem[{\citenamefont{Wang and Carter}(2000)}]{WangCarter00}
\bibinfo{author}{\bibfnamefont{Y.~A.} \bibnamefont{Wang}} \bibnamefont{and}
  \bibinfo{author}{\bibfnamefont{E.~A.} \bibnamefont{Carter}}, in
  \emph{\bibinfo{booktitle}{Theoretical Methods in Condensed Phase Chemistry}},
  edited by \bibinfo{editor}{\bibfnamefont{S.~D.} \bibnamefont{Schwartz}}
  (\bibinfo{publisher}{Kluwer}, \bibinfo{year}{2000}), pp.
  \bibinfo{pages}{117--184}.

\bibitem[{\citenamefont{Chan et~al.}(2001)\citenamefont{Chan, Cohen, and
  Handy}}]{Chan01}
\bibinfo{author}{\bibfnamefont{G.~K.} \bibnamefont{Chan}},
  \bibinfo{author}{\bibfnamefont{A.~J.} \bibnamefont{Cohen}}, \bibnamefont{and}
  \bibinfo{author}{\bibfnamefont{N.~C.} \bibnamefont{Handy}},
  \bibinfo{journal}{J. Chem. Phys.} \textbf{\bibinfo{volume}{114}},
  \bibinfo{pages}{631} (\bibinfo{year}{2001}).

\bibitem[{\citenamefont{Shah et~al.}(1994)\citenamefont{Shah, Nehete, and
  Kanhere}}]{Shah94}
\bibinfo{author}{\bibfnamefont{V.}~\bibnamefont{Shah}},
  \bibinfo{author}{\bibfnamefont{D.}~\bibnamefont{Nehete}}, \bibnamefont{and}
  \bibinfo{author}{\bibfnamefont{D.~G.} \bibnamefont{Kanhere}},
  \bibinfo{journal}{J. Phys.:Condens. Matter} \textbf{\bibinfo{volume}{6}},
  \bibinfo{pages}{10773} (\bibinfo{year}{1994}).

\bibitem[{\citenamefont{Smargiassi and Madden}(1994)}]{Smargiassi94}
\bibinfo{author}{\bibfnamefont{E.}~\bibnamefont{Smargiassi}} \bibnamefont{and}
  \bibinfo{author}{\bibfnamefont{P.~A.} \bibnamefont{Madden}},
  \bibinfo{journal}{Phys. Rev. B} \textbf{\bibinfo{volume}{49}},
  \bibinfo{pages}{5220} (\bibinfo{year}{1994}).

\bibitem[{\citenamefont{Govind et~al.}(1994)\citenamefont{Govind, Wang, and
  Guo}}]{Govind94}
\bibinfo{author}{\bibfnamefont{N.}~\bibnamefont{Govind}},
  \bibinfo{author}{\bibfnamefont{J.}~\bibnamefont{Wang}}, \bibnamefont{and}
  \bibinfo{author}{\bibfnamefont{H.}~\bibnamefont{Guo}},
  \bibinfo{journal}{Phys. Rev. B} \textbf{\bibinfo{volume}{50}},
  \bibinfo{pages}{11175} (\bibinfo{year}{1994}).

\bibitem[{\citenamefont{Blaise et~al.}(1997)\citenamefont{Blaise, Blundell, and
  Guet}}]{Blaise97}
\bibinfo{author}{\bibfnamefont{P.}~\bibnamefont{Blaise}},
  \bibinfo{author}{\bibfnamefont{S.~A.} \bibnamefont{Blundell}},
  \bibnamefont{and} \bibinfo{author}{\bibfnamefont{C.}~\bibnamefont{Guet}},
  \bibinfo{journal}{Phys. Rev. B} \textbf{\bibinfo{volume}{55}},
  \bibinfo{pages}{15856} (\bibinfo{year}{1997}).

\bibitem[{\citenamefont{Foley and Madden}(1996)}]{Foley96}
\bibinfo{author}{\bibfnamefont{M.}~\bibnamefont{Foley}} \bibnamefont{and}
  \bibinfo{author}{\bibfnamefont{P.~A.} \bibnamefont{Madden}},
  \bibinfo{journal}{Phys. Rev. B} \textbf{\bibinfo{volume}{53}},
  \bibinfo{pages}{10589} (\bibinfo{year}{1996}).

\bibitem[{\citenamefont{Domps et~al.}(1998)\citenamefont{Domps, Reinhard, and
  Suraud}}]{Domps98}
\bibinfo{author}{\bibfnamefont{A.}~\bibnamefont{Domps}},
  \bibinfo{author}{\bibfnamefont{P.~G.} \bibnamefont{Reinhard}},
  \bibnamefont{and} \bibinfo{author}{\bibfnamefont{E.}~\bibnamefont{Suraud}},
  \bibinfo{journal}{Phys. Rev. Lett.} \textbf{\bibinfo{volume}{80}},
  \bibinfo{pages}{5520} (\bibinfo{year}{1998}).

\bibitem[{\citenamefont{Watson and Carter}(2000)}]{Watson00}
\bibinfo{author}{\bibfnamefont{S.~C.} \bibnamefont{Watson}} \bibnamefont{and}
  \bibinfo{author}{\bibfnamefont{E.~A.} \bibnamefont{Carter}},
  \bibinfo{journal}{Comp. Phys. Comm.} \textbf{\bibinfo{volume}{128}},
  \bibinfo{pages}{67} (\bibinfo{year}{2000}).

\bibitem[{\citenamefont{Wang et~al.}(1995)\citenamefont{Wang, Wang, Guo, and
  Zaremba}}]{Wang95}
\bibinfo{author}{\bibfnamefont{Y.}~\bibnamefont{Wang}},
  \bibinfo{author}{\bibfnamefont{J.}~\bibnamefont{Wang}},
  \bibinfo{author}{\bibfnamefont{H.}~\bibnamefont{Guo}}, \bibnamefont{and}
  \bibinfo{author}{\bibfnamefont{E.}~\bibnamefont{Zaremba}},
  \bibinfo{journal}{Phys. Rev. B} \textbf{\bibinfo{volume}{52}},
  \bibinfo{pages}{2738} (\bibinfo{year}{1995}).

\bibitem[{\citenamefont{Sinha et~al.}(2000)\citenamefont{Sinha, Shankar, and
  Murthy}}]{Sinha00}
\bibinfo{author}{\bibfnamefont{S.}~\bibnamefont{Sinha}},
  \bibinfo{author}{\bibfnamefont{R.}~\bibnamefont{Shankar}}, \bibnamefont{and}
  \bibinfo{author}{\bibfnamefont{M.~V.~N.} \bibnamefont{Murthy}},
  \bibinfo{journal}{Phys. Rev. B} \textbf{\bibinfo{volume}{62}},
  \bibinfo{pages}{10896} (\bibinfo{year}{2000}).

\bibitem[{\citenamefont{Shi and Xie}(2002)}]{Shi02}
\bibinfo{author}{\bibfnamefont{J.}~\bibnamefont{Shi}} \bibnamefont{and}
  \bibinfo{author}{\bibfnamefont{X.~C.} \bibnamefont{Xie}},
  \bibinfo{journal}{Phys. Rev. Lett.} \textbf{\bibinfo{volume}{88}},
  \bibinfo{pages}{086401} (\bibinfo{year}{2002}).

\bibitem[{\citenamefont{Ullmo and Baranger}(2001)}]{Ullmo01b}
\bibinfo{author}{\bibfnamefont{D.}~\bibnamefont{Ullmo}} \bibnamefont{and}
  \bibinfo{author}{\bibfnamefont{H.~U.} \bibnamefont{Baranger}},
  \bibinfo{journal}{Phys. Rev. B} \textbf{\bibinfo{volume}{64}},
  \bibinfo{pages}{245324} (\bibinfo{year}{2001}).

\bibitem[{\citenamefont{Strutinsky}(1968)}]{Strutinsky68}
\bibinfo{author}{\bibfnamefont{V.~M.} \bibnamefont{Strutinsky}},
  \bibinfo{journal}{Nucl. Phys. A} \textbf{\bibinfo{volume}{122}},
  \bibinfo{pages}{1} (\bibinfo{year}{1968}).

\bibitem[{\citenamefont{Ullmo et~al.}(2001)\citenamefont{Ullmo, Nagano,
  Tomosovic, and Baranger}}]{Ullmo01a}
\bibinfo{author}{\bibfnamefont{D.}~\bibnamefont{Ullmo}},
  \bibinfo{author}{\bibfnamefont{T.}~\bibnamefont{Nagano}},
  \bibinfo{author}{\bibfnamefont{S.}~\bibnamefont{Tomosovic}},
  \bibnamefont{and} \bibinfo{author}{\bibfnamefont{H.~U.}
  \bibnamefont{Baranger}}, \bibinfo{journal}{Phys. Rev. B}
  \textbf{\bibinfo{volume}{63}}, \bibinfo{pages}{125339}
  (\bibinfo{year}{2001}).

\bibitem[{\citenamefont{Payne et~al.}(1992)\citenamefont{Payne, Teter, Allan,
  Arias, and Joannopoulos}}]{Payne92RMP}
\bibinfo{author}{\bibfnamefont{M.~C.} \bibnamefont{Payne}},
  \bibinfo{author}{\bibfnamefont{M.~P.} \bibnamefont{Teter}},
  \bibinfo{author}{\bibfnamefont{D.~C.} \bibnamefont{Allan}},
  \bibinfo{author}{\bibfnamefont{T.~A.} \bibnamefont{Arias}}, \bibnamefont{and}
  \bibinfo{author}{\bibfnamefont{J.~D.} \bibnamefont{Joannopoulos}},
  \bibinfo{journal}{Rev. Mod. Phys.} \textbf{\bibinfo{volume}{64}},
  \bibinfo{pages}{1045} (\bibinfo{year}{1992}), \bibinfo{note}{and references
  therein}.

\bibitem[{\citenamefont{Kresse and Furthm{\"u}ller}(1996)}]{Kresse96}
\bibinfo{author}{\bibfnamefont{G.}~\bibnamefont{Kresse}} \bibnamefont{and}
  \bibinfo{author}{\bibfnamefont{J.}~\bibnamefont{Furthm{\"u}ller}},
  \bibinfo{journal}{Phys. Rev. B} \textbf{\bibinfo{volume}{54}},
  \bibinfo{pages}{11169} (\bibinfo{year}{1996}), \bibinfo{note}{and references
  therein}.

\bibitem[{\citenamefont{Press et~al.}(1989)\citenamefont{Press, Flannery,
  Teukolsky, and Vetterlin}}]{NumRecip}
\bibinfo{author}{\bibfnamefont{W.~H.} \bibnamefont{Press}},
  \bibinfo{author}{\bibfnamefont{B.~P.} \bibnamefont{Flannery}},
  \bibinfo{author}{\bibfnamefont{S.~A.} \bibnamefont{Teukolsky}},
  \bibnamefont{and} \bibinfo{author}{\bibfnamefont{W.~T.}
  \bibnamefont{Vetterlin}}, \emph{\bibinfo{title}{Numerical Recipes: The Art of
  Scientific Computing}} (\bibinfo{publisher}{Cambridge University},
  \bibinfo{address}{Cambridge, England}, \bibinfo{year}{1989}).

\bibitem[{\citenamefont{Teter et~al.}(1989)\citenamefont{Teter, Payne, and
  Allan}}]{Teter89}
\bibinfo{author}{\bibfnamefont{M.~P.} \bibnamefont{Teter}},
  \bibinfo{author}{\bibfnamefont{M.~C.} \bibnamefont{Payne}}, \bibnamefont{and}
  \bibinfo{author}{\bibfnamefont{D.~C.} \bibnamefont{Allan}},
  \bibinfo{journal}{Phys. Rev. B} \textbf{\bibinfo{volume}{40}},
  \bibinfo{pages}{12255} (\bibinfo{year}{1989}).

\bibitem[{\citenamefont{Jiang et~al.}(2003{\natexlab{a}})\citenamefont{Jiang,
  Baranger, and Yang}}]{Jiang03b}
\bibinfo{author}{\bibfnamefont{H.}~\bibnamefont{Jiang}},
  \bibinfo{author}{\bibfnamefont{H.~U.} \bibnamefont{Baranger}},
  \bibnamefont{and} \bibinfo{author}{\bibfnamefont{W.}~\bibnamefont{Yang}},
  \bibinfo{journal}{Phys. Rev. B} \textbf{\bibinfo{volume}{68}},
  \bibinfo{pages}{165337} (\bibinfo{year}{2003}{\natexlab{a}}).

\bibitem[{\citenamefont{Oliver and Perdew}(1979)}]{Oliver79}
\bibinfo{author}{\bibfnamefont{G.~L.} \bibnamefont{Oliver}} \bibnamefont{and}
  \bibinfo{author}{\bibfnamefont{J.~P.} \bibnamefont{Perdew}},
  \bibinfo{journal}{Phys. Rev. A} \textbf{\bibinfo{volume}{20}},
  \bibinfo{pages}{397} (\bibinfo{year}{1979}).

\bibitem[{\citenamefont{Govind}(1995)}]{Govind95}
\bibinfo{author}{\bibfnamefont{N.}~\bibnamefont{Govind}}, Ph.D. thesis,
  \bibinfo{school}{McGill University, Canada} (\bibinfo{year}{1995}).

\bibitem[{\citenamefont{Jiang et~al.}(2003{\natexlab{b}})\citenamefont{Jiang,
  Baranger, and Yang}}]{Jiang03}
\bibinfo{author}{\bibfnamefont{H.}~\bibnamefont{Jiang}},
  \bibinfo{author}{\bibfnamefont{H.~U.} \bibnamefont{Baranger}},
  \bibnamefont{and} \bibinfo{author}{\bibfnamefont{W.}~\bibnamefont{Yang}},
  \bibinfo{journal}{Phys. Rev. Lett.} \textbf{\bibinfo{volume}{90}},
  \bibinfo{pages}{026806} (\bibinfo{year}{2003}{\natexlab{b}}).

\bibitem[{\citenamefont{Bertsekas}(1999)}]{Bertsekas99}
\bibinfo{author}{\bibfnamefont{D.~P.} \bibnamefont{Bertsekas}},
  \emph{\bibinfo{title}{Nonlinear programming}} (\bibinfo{publisher}{Athena},
  \bibinfo{address}{Belmont, Massachusetts}, \bibinfo{year}{1999}).

\bibitem[{\citenamefont{Martyna and Tuckerman}(1999)}]{Martyna99}
\bibinfo{author}{\bibfnamefont{G.~J.} \bibnamefont{Martyna}} \bibnamefont{and}
  \bibinfo{author}{\bibfnamefont{M.~E.} \bibnamefont{Tuckerman}},
  \bibinfo{journal}{J. Chem. Phys.} \textbf{\bibinfo{volume}{110}},
  \bibinfo{pages}{2810} (\bibinfo{year}{1999}).

\bibitem[{\citenamefont{Tanatar and Ceperley}(1989)}]{Tanatar89}
\bibinfo{author}{\bibfnamefont{B.}~\bibnamefont{Tanatar}} \bibnamefont{and}
  \bibinfo{author}{\bibfnamefont{D.~M.} \bibnamefont{Ceperley}},
  \bibinfo{journal}{Phys. Rev. B} \textbf{\bibinfo{volume}{39}},
  \bibinfo{pages}{5005} (\bibinfo{year}{1989}).

\bibitem[{\citenamefont{Vosko et~al.}(1980)\citenamefont{Vosko, Wilk, and
  Nusair}}]{Vosko80}
\bibinfo{author}{\bibfnamefont{S.~H.} \bibnamefont{Vosko}},
  \bibinfo{author}{\bibfnamefont{L.}~\bibnamefont{Wilk}}, \bibnamefont{and}
  \bibinfo{author}{\bibfnamefont{M.}~\bibnamefont{Nusair}},
  \bibinfo{journal}{Can. J. Phys.} \textbf{\bibinfo{volume}{58}},
  \bibinfo{pages}{100} (\bibinfo{year}{1980}).

\bibitem[{\citenamefont{Bohigas et~al.}(1993)\citenamefont{Bohigas, Tomsovic,
  and Ullmo}}]{Bohigas93}
\bibinfo{author}{\bibfnamefont{O.}~\bibnamefont{Bohigas}},
  \bibinfo{author}{\bibfnamefont{S.}~\bibnamefont{Tomsovic}}, \bibnamefont{and}
  \bibinfo{author}{\bibfnamefont{D.}~\bibnamefont{Ullmo}},
  \bibinfo{journal}{Phys. Rep.} \textbf{\bibinfo{volume}{223}},
  \bibinfo{pages}{43} (\bibinfo{year}{1993}).

\bibitem[{\citenamefont{Bachelet et~al.}(1982)\citenamefont{Bachelet, Hamann,
  and Schl\"uter}}]{Bachelet82}
\bibinfo{author}{\bibfnamefont{G.~B.} \bibnamefont{Bachelet}},
  \bibinfo{author}{\bibfnamefont{D.~R.} \bibnamefont{Hamann}},
  \bibnamefont{and}
  \bibinfo{author}{\bibfnamefont{M.}~\bibnamefont{Schl\"uter}},
  \bibinfo{journal}{Phys. Rev. B} \textbf{\bibinfo{volume}{26}},
  \bibinfo{pages}{4199} (\bibinfo{year}{1982}).

\end{thebibliography}

\end{document}